# Magnetic effects at the interface between nonmagnetic oxides

A. Brinkman[1], M. Huijben[1,*], M. van Zalk[1], J. Huijben[1], U. Zeitler[2], J.C. Maan[2], W.G. van der Wiel[3], G. Rijnders[1], D.H.A. Blank[1], and H. Hilgenkamp[1]

[1]*Faculty of Science and Technology and MESA⁺ Institute for Nanotechnology, University of Twente, The Netherlands*

[2]*High Field Magnet Laboratory, Institute for Molecules and Materials, Radboud University Nijmegen, The Netherlands*

[3]*Strategic Research Orientation NanoElectronics, MESA⁺ Institute for Nanotechnology, University of Twente, The Netherlands*

**The electronic reconstruction at the interface between two insulating oxides can give rise to a highly-conductive interface[1,2]. In analogy to this remarkable interface-induced conductivity we show how, additionally, magnetism can be induced at the interface between the otherwise nonmagnetic insulating perovskites $SrTiO_3$ and $LaAlO_3$. A large negative magnetoresistance of the interface is found, together with a logarithmic temperature dependence of the sheet resistance. At low temperatures, the sheet resistance reveals magnetic hysteresis. Magnetic ordering is a key issue in solid-state science and its underlying mechanisms are still the subject of intense research. In particular, the interplay between localized magnetic moments and the spin of itinerant conduction electrons in a solid gives rise to intriguing many-body effects such as Ruderman-Kittel-Kasuya-Yosida (RKKY) interactions[3], the Kondo effect[4], and carrier-induced ferromagnetism in diluted magnetic semiconductors[5]. The conducting oxide interface now provides a versatile**

[*] Present address: Physics Department, University of California, Berkeley, USA

**system to induce and manipulate magnetic moments in otherwise nonmagnetic materials.**

The discovery[1,2] of conduction caused by electronic reconstruction at oxide interfaces has attracted a lot of interest[6-13]. The general structure of the perovskites $ABO_3$, with A and B being cations, can be regarded as a stack of alternating sub-unit-cell layers of AO and $BO_2$. A heterointerface introduces polarity discontinuities when both elements A and B on either side of the interface have different valence states. Recently, Ohtomo and Hwang[2] found different electronic behaviour for thin $LaAlO_3$ films on either SrO or $TiO_2$ terminated $SrTiO_3$ substrates, the former interface being insulating and the latter interface being an *n*-type conductor. Assuming the formal valence states, the polarity discontinuities can be described by either the $(SrO)^0$-$(AlO_2)^-$ or $(TiO_2)^0$-$(LaO)^+$ sequence[6]. Similar behaviour was found for the $LaTiO_3/SrTiO_3$ interface[1], resulting in a surplus of half an electron per unit cell at the $TiO_2$-LaO interface, entering the otherwise empty conduction band. In addition to charge degrees of freedom, also an induction of spin states at interfaces can be expected. Following this scenario, a ferromagnetic alignment of the induced electron spins within the Ti-3*d* conduction band was theoretically predicted for the $LaTiO_3/SrTiO_3$ interface[14] as well as for the $LaAlO_3/SrTiO_3$ interface[15].

To appreciate the novelty of this predicted interface magnetism, the comparison can be made to doped $SrTiO_3$. $SrTiO_3$ is a band insulator with a bandgap of about 3.2 eV. Doping with $La^{3+}$ on the $Sr^{2+}$ site introduces itinerant charge carriers in the conduction band. This doping gives rise to conductivity, but not to magnetism, since the Ti remains in its nominal valence 4+ state, with itinerant electrons in hybridized *spd* bands rather than localized in the 3*d* shell, i.e. $3d^0$. Only when almost all Sr is replaced by La, the Ti obtains a formal valence of 3+ with a spin-½ local moment[16], i.e. $3d^1$. The limiting case is $LaTiO_3$, which is an antiferromagnetic Mott insulator. The search for ferromagnetism in these systems is ongoing, but has only been successful when large amounts of magnetic elements like several percents of Co or at least 10% of Cr on the Ti site are added[17,18]. In this case, the local moments, such as the spin-3/2 $Cr^{3+}$ $3d^3$ moment, give rise to magnetic effects on the transport properties when they are coupled to doping-induced itinerant charge carriers[18].

In order to realize the $SrTiO_3/LaAlO_3$ interface, we used a $SrTiO_3$ (001) substrate that was $TiO_2$-terminated by a buffered-HF and annealing treatment[19]. An atomically smooth surface with clear unit-cell-height steps was observed with atomic force microscopy (AFM). On top of that, a $LaAlO_3$ film was grown at 850 ºC in a wide range of partial oxygen pressures by pulsed laser deposition using a single crystal $LaAlO_3$ target. The growth was monitored by in-situ reflective high energy electron diffraction (RHEED)[20]. The observed intensity oscillations indicated clearly a layer-by-layer growth mode with a very smooth surface and without any island growth. The thickness of the $LaAlO_3$ layer was 26 unit cells, approximately 10 nm, as determined from the RHEED oscillations. After growth, the sample was slowly cooled down to room temperature in oxygen at the deposition pressure. A smooth surface morphology and the correct crystal structure of the final sample were confirmed by AFM and x-ray diffraction. X-ray photoelectron spectroscopy (XPS) at a treated $SrTiO_3$ substrate and a deposited film did not show any trace of magnetic elements such as Fe, Co, Cr, and Mn within a resolution of 0.1 %. Four ohmic Al contacts were wire-bonded to the corners of the samples, allowing to contact the interface and measure the transport properties without structuring the sample, which could induce damage to the interface.

The sheet resistance, $R_S$, as measured in a Van-der-Pauw geometry is shown in Fig. 1a for different $n$-type $SrTiO_3/LaAlO_3$ interfaces, grown at various partial oxygen pressures (while still maintaining a two-dimensional growth mode), from which the presence of a conducting interface is evident. The temperature dependencies of the corresponding Hall coefficients, $-1/eR_H$, are shown and described as supplementary information. Several groups have found that oxygen vacancies give rise to conduction in oxide interfaces when the oxide layers are deposited at low oxygen pressure[21,22], which also becomes clear from the dependence of $R_S$ on the deposition conditions. From here on we will focus on the transport properties of the samples deposited at 1.0 and 2.5 × $10^{-3}$ mbar in which the influence of oxygen vacancies is the lowest (see the supplementary information).

In order to investigate the magnetic properties of the interface, the magnetic field dependence of $R_S$ is measured. Figure 2a shows the measured $R_S$ as function of magnetic field at different temperatures. We define the magnetoresistance as the change in resistance relative to the zero-field resistance, $[R_S(H) - R_S(0)] / R_S(0)$. A large negative magnetoresistance effect is observed in both samples on the order of 30% over a magnetic field range of 30 T.

The magnetoresistance of the conducting interface is independent of the orientation of the magnetic field relative to the interface, which shows that the large magnetoresistance is related to spin physics and not to orbital effects (such as weak localization). We therefore suppose that the observed behaviour must be ascribed to spin scattering of conduction electrons off localized magnetic moments at the interface, as we will substantiate below.

Most theoretical treatments of spin scattering in metals are elaborations of the s-d model, the non-degenerate Anderson model, or the Kondo model, all of which describe the interaction between itinerant charge carriers and localized magnetic moments[23]. The scattering cross-section of conduction electrons off localized magnetic moments depends on the relative spin orientation. Under an applied magnetic field, spin-flip scattering off localized moments is suppressed at the Fermi level, because of the finite

Zeeman splitting between the spin-up and spin-down levels of the localized magnetic moments. The Zeeman splitting thus turns the spin scattering into an inelastic process, requiring energy exchange with the environment. The large negative magnetoresistance (up to 70%) observed in (Sr,La)TiO$_3$ alloyed with 20% of Cr has been explained by an enhanced coherent motion of the conduction electrons due to ferromagnetically aligned spins[17].

In the case of the *n*-type interface, the temperature dependence of the sheet resistance helps to further understand the nature of the scattering. The temperature dependence of the sheet resistance is found to be logarithmic over one decade (~5-50K, see Fig. 1b). The sheet resistance can be described by $R_S = a \ln(T/T_{eff}) + bT^2 + cT^5$, where $T_{eff} \sim 70$ K is an effective crossover temperature scale, and where the $T^2$ and $T^5$ terms are suggestive of electron-electron and electron-phonon scattering, relevant at higher temperatures. Saturation of the logarithmic term is observed below ~5 K. Additionally, we observe a voltage-induced resistance suppression at low temperatures, as shown in the inset of Fig. 1b, in which a four-point differential resistance measurement is shown at the temperatures indicated in Fig. 1b.

An explanation, although still suggestive at the moment, for the observed logarithmic temperature dependence of the sheet resistance is the Kondo effect[4], which describes the interplay between localized magnetic moments and mobile charge carriers[4,24,25]. $T_{eff}$ can then be interpreted as the Kondo temperature. The Kondo temperature in this definition depends on other temperature-independent scatter terms, which in the present case, however, are small, as can be seen from the depth of the minimum in Fig. 1b. The energy scale $k_B T_{eff}$ is of the same order of magnitude as the observed energy scale $g\mu_B B$ over which magnetoresistance effects occur. The suppression of the Kondo effect at finite bias voltage in single-impurity (quantum dot) systems[25,26] arises from the fact that the Kondo resonance is pinned to the Fermi level and that an excursion of the conduction electrons from the Fermi level (due to a finite bias) decreases the Kondo scattering. For an ensemble of impurities it is much more difficult to apply a well-defined bias voltage, and in the present case, the voltage drop per unit cell (in our case of the order of microvolts) is only very small. Therefore, the interpretation of our experimental observations in terms of the Kondo effect is, as said, still speculative.

Magnetoresistance in spin scattering models in general is proportional to the square of the global magnetization[23], $[R_S(H) - R_S(0)] / R_S(0) \sim -M^2(H)$, which allows to extract a measure of $M(H)$ from the magnetoresistance data of Fig. 2a. The derived magnetization is depicted in Fig. 2b and is linear at low magnetic fields, while saturating at high fields. The kink at small fields is an artifact of the model that does not take into account the small positive contribution to the magnetoresistance as visible in Fig. 2a. The nature of the positive contribution needs to be further investigated, but could be related to two-band conduction effects (see the supplementary information). The low-field susceptibility $\chi = dM/dH$ for the different temperatures is plotted in the inset of Fig. 2b, from which a Curie-Weiss dependence $\chi = C/(T+\theta)$ is found, where $C$ is a constant and $\theta$ an offset temperature, characteristic for antiferromagnetic coupling at an energy scale of $k_B\theta = 40$ $\mu$eV. The competition between this energy scale and a possible Kondo singlet ground state depends on the development of the exchange interactions as function of temperature and density of states[27]. A locally *anti*ferromagnetic interaction between mobile and localized spins could in principle cause a *ferro*magnetic alignment of the localized spins, as in the RKKY model of iterant ferromagnetism[3].

At 0.3 K, hysteresis appears in the sheet resistance, as shown in Fig. 3a. Again, the magnetoresistance is independent of field orientation. Magnetoresistance hysteresis is usually indicative of ferromagnetic domain formation in which domains change polarity above a certain coercive field. Domain formation typically creates a remanence in the signal when crossing zero-field, providing a butterfly shape of the magnetoresistance curve. In the present case, the shape cannot completely be explained by such an overshoot of the signal, but an additional suppression around zero-field seems to occur, which could suggest additional spin/domain reorientation effects, such as observed in granular and spin-valve giant magnetoresistance systems, and the Kondo effect in quantum dots in the presence of ferromagnetism[28]. The presence of a small positive contribution to the magnetoresistance of Fig. 2a might be related to the shape of the hysteresis curve. We observe that the domain formation in $n$-type $SrTiO_3$-$LaAlO_3$ conducting interfaces is a rather slow process. In Fig. 3b it is shown how the sheet resistance exponentially approaches the saturation value at a constant field of 4 T, with a time constant of about 10 s. The long time constant has been observed in mixed valence manganites as well, where phase separation between charge-ordered and ferromagnetic phases gives rise to the formation, expansion and destruction of domains[29].

The explanation for the temperature dependence of $R_S$, the negative magnetoresistance, and the observed hysteresis in $n$-type $SrTiO_3$-$LaAlO_3$ conducting interfaces in terms of the magnetic scattering centres requires the presence of magnetic moments. Magnetic effects from carrier doping or oxygen vacancies in $Sr_xLa_{1-x}TiO_3$ systems have been sought for, but have never been observed[17]. Although very low impurity concentrations are known to give rise to the Kondo effect, for the observed strong $R_S(T)$ minimum and large magnetoresistance of 30%, orders of magnitude more impurities (on the order of 10%, as determined from comparable experimental studies[17,18]) are required than the upper limit that was determined from the XPS analysis on our samples (maximally 0.1%). The necessary magnetic moments can only arise from the polar discontinuity, as schematically illustrated in Fig. 1c. The interface between two non-magnetic materials has itself become magnetic.

Our present finding of a large magnetoresistance and magnetic hysteresis at the interface between $TiO_2$-terminated $SrTiO_3$ and $LaAlO_3$, also sheds light on the influence of oxygen vacancies in related interface structures. Importantly, with the present observation of magnetism, we have shown that the polar discontinuity is in fact present and that the interface conductivity cannot solely be attributed to oxygen vacancies in the considered range of deposition pressures.

The conducting oxide interface thus forms an intriguing system for studying fundamental magnetic interactions in solid-state systems and possibly opens up the way for carrier-controlled ferromagnetism in all-oxide devices.

**Acknowledgements** This work is part of the research program of the Foundation for Fundamental Research on Matter [FOM, financially supported by the Netherlands Organization for Scientific Research (NWO)]. A.B., D.B., H.H., G.R. and W.v.d.W. acknowledge additional support from NWO. The authors acknowledge J. Mannhart, A.J. Millis, R. Pentcheva, and D. Winkler for useful discussions.

**Author Information** The authors declare no competing financial interests. Correspondence and requests for materials should be addressed to A.B. (a.brinkman@utwente.nl).

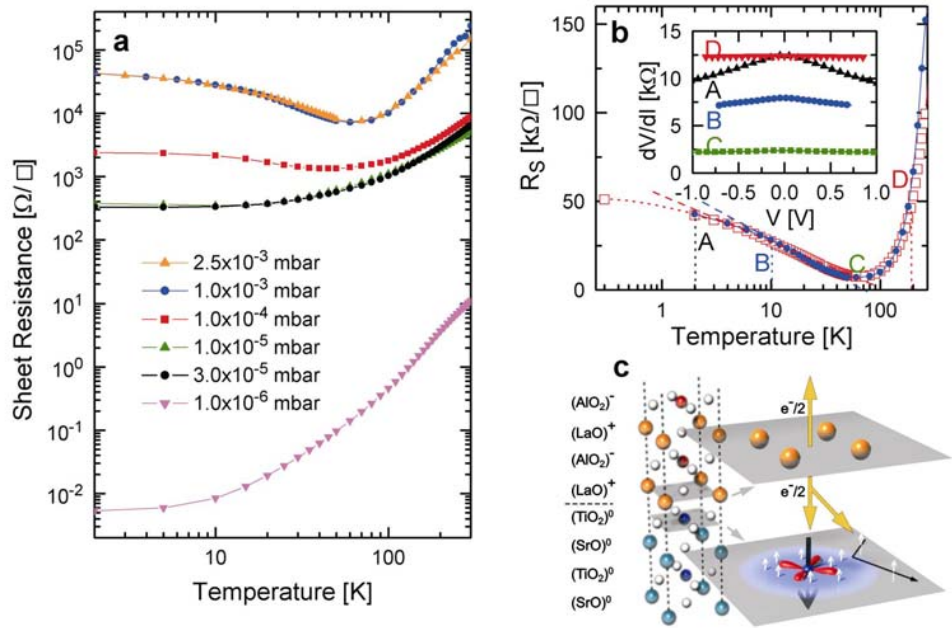

**Figure 1 | Sheet resistance of *n*-type SrTiO$_3$-LaAlO$_3$ interfaces. a,** Temperature dependence of the sheet resistance $R_S$ for *n*-type SrTiO$_3$-LaAlO$_3$ conducting interfaces, grown at various partial oxygen pressures. **b,** Temperature dependence of the sheet resistance $R_S$ for two conducting interfaces, grown respectively at a partial oxygen pressure of 2.5 × 10$^{-3}$ mbar (□) and 1.0 × 10$^{-3}$ mbar (●). The low temperature logarithmic dependencies are indicated by dashed lines. Inset: four-point differential resistance d$V$/d$I$ as function of applied voltage, at a constant temperature of 2.0 K (A), 10.0 K (B), 50.0 K (C) and 180.0 K (D). **c,** Schematic representation of the electron transfer from the LaO layer into the TiO$_2$ layer. The electrons either form localized 3$d$ magnetic moments on the Ti site, or conduction electrons that can scatter off the Kondo cloud surrounding the localized moments.

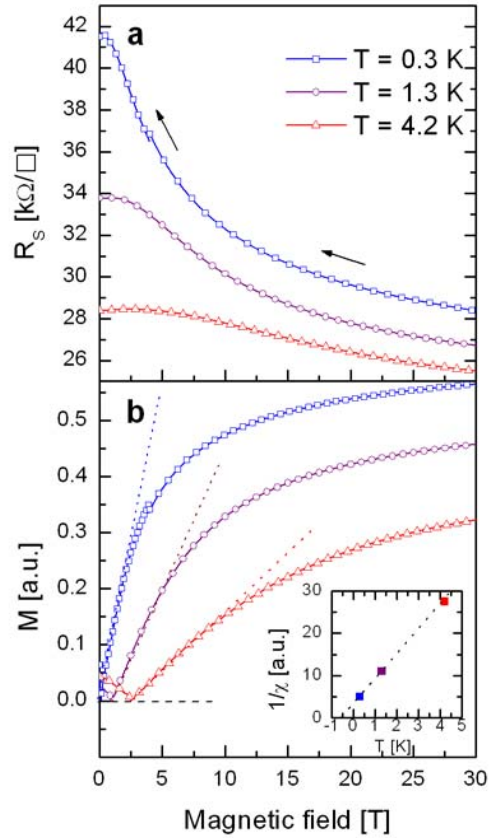

**Figure 2 | Large negative magnetoresistance. a,** Sheet resistance $R_S$ of an *n*-type SrTiO$_3$-LaAlO$_3$ conducting interface, grown at $1.0 \times 10^{-3}$ mbar, under applied magnetic field perpendicular to the interface at 0.3 K, 1.3 K and 4.2 K. The magnetic field sweep direction is indicated by arrows. **b,** Magnetization in units of $g\mu_B/\pi$ as function of the applied field, as inferred from a quadratic dependence of the magnetoresistance in Fig. 2a on magnetization. The susceptibility is derived from the linear slope at low fields. Inset: low-field inverse susceptibility as function of temperature, fitted by the Curie-Weiss law $\chi = C/(T+\theta)$ with $\theta = 0.5$ K.

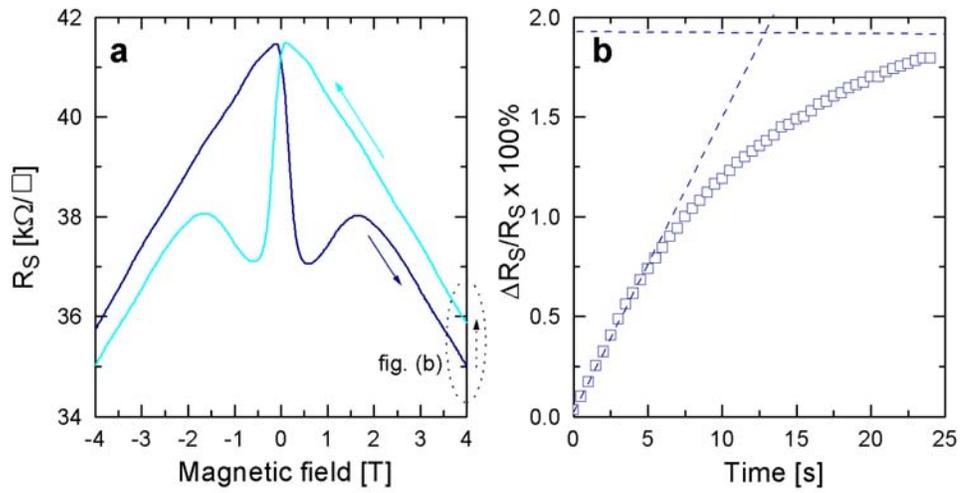

**Figure 3 | Magnetic hysteresis. a,** Sheet resistance at 0.3 K of an *n*-type SrTiO$_3$-LaAlO$_3$ conducting interface, grown at 1.0 × 10$^{-3}$ mbar. Arrows indicate the direction of the measurements (at a rate of 30 mT/s). **b,** Relative sheet resistance saturation at a constant magnetic field of 4 T, as indicated in a.

**Supplementary information: Hall data and two-band conduction**

The conduction of the LaAlO$_3$/SrTiO$_3$ sample that is deposited at $10^{-6}$ mbar is completely dominated by oxygen vacancies, with a carrier density $n_{ox}$ of the order of $10^{17}$ cm$^{-2}$, as can be directly obtained from the Hall data of Supplementary Figure 1. The mobility of these carriers increases at low temperatures to about $10^4$ cm$^2$V$^{-1}$s$^{-1}$. With this many oxygen vacancies there is no need for a polar discontinuity at the interface [1,2]. The number of oxygen vacancies can be reduced by depositing the samples at higher oxygen deposition pressures. When $n_{ox}$ drops well below half an electron per unit cell (of the order of $10^{14}$ cm$^{-2}$), the polar discontinuity is sustained and interface induced carriers $n_{int}$ will contribute to conduction. When two contributions to conduction (index 1 and 2) are taken into account, the equations for sheet- and Hall resistance are generally written as

$$R_H = \frac{n_1\mu_1^2 + n_2\mu_2^2}{e(n_1\mu_1 + n_2\mu_2)^2},$$
$$R_S = e(n_1\mu_1 + n_2\mu_2)^{-1}$$

The Hall and sheet resistance data of all our samples can be fitted with this two-band model [3], where the bands 1 and 2 are likely to be identified as arising from the oxygen vacancies and the interface respectively.

For the 1.0 and 2.5×10$^{-3}$ mbar samples, the sheet resistance is found to be determined by the interface induced carriers ($n_{int}\mu_{int}$ being larger than $n_{ox}\mu_{ox}$) despite the fact that $\mu_{ox}$ is much larger than $\mu_{int}$ (the latter being of the order of 1 cm$^2$V$^{-1}$s$^{-1}$), which is of relevance to the interpretation of the magnetoresistance data of this manuscript.

However, in the nominator of the expression for $R_H$, the squared mobility enters, making the oxygen vacancies important in the interpretation of $R_H$. It is found that $n_{ox}$ at high temperatures is about $10^{11}$ cm$^{-2}$, and that this density is strongly reduced at low temperatures. This gapping of the carriers is expected since the impurity band becomes so narrow (few carriers) that it no longer intersects the Fermi level [4]. The low temperature carrier freeze-out of $n_{ox}$ gives rise to the observed upturn in $1/eR_H$ in Supplementary Figure 1.

Concluding, in the samples deposited at relatively high oxygen pressure, the sheet resistance is determined by the interface induced carriers. The influence of a small amount of oxygen vacancies only becomes apparent in the measured Hall resistance. Additionally, for two contributing conduction bands, a small positive contribution to magnetoresistance is to be expected.

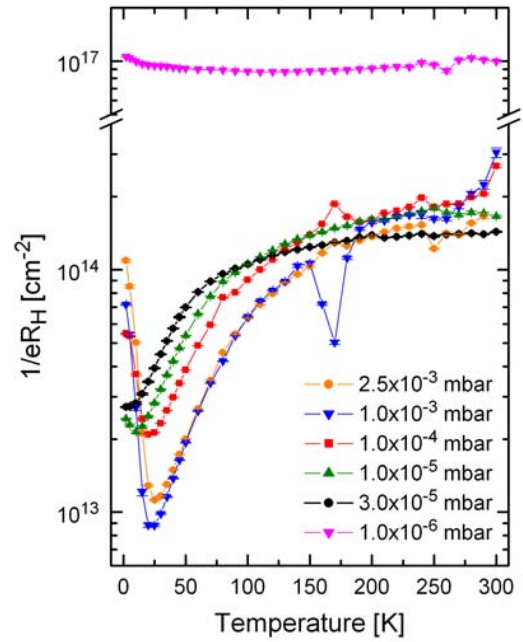

**Supplementary Figure 1 | Hall coefficient.** Temperature dependence of the inverse Hall resistance of *n*-type $SrTiO_3$-$LaAlO_3$ conducting interfaces, grown at various partial oxygen deposition pressures.